\documentstyle[aps,prl,amstex,graphicx]{revtex}
\begin{document}
\draft 
\title{Critical State Flux Penetration and Linear Microwave Vortex
  Response in ${\rm Y_1Ba_2Cu_3O_{7-\delta}}$ Films}
\author{Balam A. Willemsen$^{1,2}$, J. S. Derov$^2$ and S.Sridhar$^1$}
\address{$^1$ Physics Department, Northeastern University, Boston, MA 02115\\
  $^2$ Rome Laboratory, Hanscom AFB, Bedford, MA 01731} \date{\today}
\maketitle

\begin{abstract}
  The vortex contribution to the dc field ($H$) dependent microwave
  surface impedance $\tilde{Z}_s=R_s+iX_s$ of ${\rm
    {Y_1Ba_2Cu_3O_{7-\delta }}}$ thin films was measured using
  suspended patterned resonators. $\tilde{Z}_s(H)$ is shown to be a
  direct measure of the flux density $B(H)$ enabling a very precise
  test of models of flux penetration. Three regimes of field-dependent
  behavior were observed: (1) Initial flux penetration occurs on very
  low field scales $H_i(4.2K)\sim 100\,{\rm{Oe}}$, (2) At moderate
  fields the flux penetration into the virgin state is in excellent
  agreement with calculations based upon the {\em field-induced} Bean
  critical state for thin film geometry, parametrized by a field scale
  $H_s(4.2K)\sim J_cd\sim 0.5\,T$, (3) for very high fields $H\gg
  H_s$, the flux density is uniform and the measurements enable direct
  determination of vortex parameters such as pinning force constants
  $\alpha_p$ and vortex viscosity $\eta $. However hysteresis loops
  are in disagreement with the thin film Bean model, and instead are
  governed by the low field scale $H_i$, rather than by $H_s$.  Geometric
  barriers are insufficient to account for the observed results.
\end{abstract}


\preprint{Critical96 0.5}

\section{Introduction}

The dynamics of vortices is important over a wide spectrum of time-scales
(or frequencies). On very long time scales the problem is relevant to
dissipation at or near the critical current. On the other end of the
spectrum, at short time scales or high frequencies, an understanding of
vortex dynamics is important both for fundamental understanding and for
practical applications of microwave devices.

Yet another need is to understand flux profiles and field penetration
into thin films. Recently there have been several calculations based
upon critical state models which have yielded field and current
distributions in thin films for magnetic fields applied perpendicular
to the films, and have shown that there are qualitative differences
from bulk geometries. While some of these results have been tested in
actual experiments\cite{AForkl93a}, not all aspects of these models
are fully understood. Most experimental studies of flux penetration
and the critical state have involved measurements of magnetic
properties such as the dc magnetization, and there are few
investigations of the dynamic properties.

This paper reports on a novel approach, utilizing microwave surface
impedance measurements in patterned microstrip resonators, to understanding
vortex dynamics and flux penetration. Specially designed microstrip
resonators patterned from thin films are used as high $Q$ structures to
probe vortex response at microwave frequencies in the presence of {dc}
magnetic fields. The high sensitivity of these structures enables
measurement of the vortex response in fields from $<0.1{\rm{Oe}}$ to several T.
The complex microwave surface impedance $\tilde{Z}_s=R_s+iX_s$ was measured
as functions of magnetic field $H$, temperature $T$ and frequency $\omega $
(in the range $1$ to $20$~GHz). $\tilde{Z}_s$ is a measure of the total flux
in the sample, and hence can lead to precise tests of flux penetration in
thin films. In addition the results also yield information on vortex
parameters such as pinning force constants and viscosity.

In this paper we focus on results for the changes in surface impedance
induced in  ${\rm Y_1Ba_2Cu_3O_{7-\delta}}$ thin films as the externally applied magnetic field
is slowly increased from zero, as well as the hysteresis loops which
ensue when the field is subsequently reduced. The response clearly
shows evidence of two relevant field scales. As the field is increased
an initial linear rise gives way at $H_i$ to a super-linear behavior
followed above $H_s$ by a linear rise which persists to the highest
fields available. While virgin response for $H>H_i$ and the magnitude
of $H_s$ appear to be consistent with recent analytical expressions
for thin strips in perpendicular fields, the hysteretic response is
not so easily understood. The critical state model predicts a
counter-clockwise hysteresis loop governed by a field scale of
order $H_s$, while the experiments observe a clockwise hysteresis
loop governed by the much smaller field scale $H_i$. Geometrical
barriers are insufficient to explain the observed hysteresis loops.
The results suggest that pinning at the film edges may be weaker than
in the bulk.

\section{Experimental Details}

The experiments were designed specifically to enable high sensitivity
measurements in dc magnetic fields from $<1{\rm{Oe}}$ to several T. The essential
element is the suspended line resonator housed in a Cu package. Microwaves
are coupled in inductively, by means of loops at the ends of coaxial lines.
Similar resonators were used in studies of high field\cite{BAWillemsen94b}
and non-linear\cite{BAWillemsen95a} response, which have already been
published. We note several key features of the experiment:

\begin{itemize}
\item  The absence of superconducting ground planes (such as would be
present in a traditional stripline or microstrip resonator). This allows us
to observe effects even at very low applied magnetic fields.

\item  The presence of several modes enables measurements at several
frequencies in the $1-20$ GHz range.

\item  The ability to carry out both linear and non-linear response
experiments independently, and also combined rf+dc field experiments where
the rf and dc field scales are comparable.

\item  Variable coupling at all temperatures  enabling weak coupling over a broad range of resonator $Q$ 's.
\end{itemize}

Experiments were initially carried out in fields of
{${\text{1~mT}}<\mu _0{H}<{\text{4~T}}$ } in a Janis SuperVaritemp
Helium flow cryostat. The magnetic fields were generated using a NbTi
coil (capable of fields up to $6$~T) which is permanently mounted in
the cryostat. Current was supplied to the solenoid by a Cryomagnetics
IPS-50 Magnet Power Supply, under computer control via RS-232 using a
Cryomagnetics CIM-16. Magnet calibration runs using a LakeShore 450
Gaussmeter and a 5~ft. cryogenic axial Hall probe showed that residual
fields of the order of ${\text{10}}-{\text{100~{\rm{Oe}}/A}}$ may be
present at the center of the superconducting NbTi magnet once it had
been energized. For experiments at true zero applied static field, and
for greater sensitivity at low fields the probe was adapted to fit in
a similar flow cryostat built by Cryo Industries of America. In this
series of low-field experiments, the magnetic fields were supplied by a liquid
${\rm N_2}$ cooled Cu solenoid capable of fields up to {400~{\rm{Oe}}}
when supplied using an HP~6024A power supply. The CIM-16 interface
module was also used here to control the supply current remotely via
RS-232. Current (and hence field) reversal was accomplished by
manually interchanging the magnet leads at zero current. The
field--to--current ratio was determined using a LakeShore 450
Gaussmeter and an axial probe to be ${\text{106}}\pm
{\text{1~{\rm{Oe}}/A}}$

A number of additional features were incorporated into the probe in
order to increase the system's sensitivity at low magnetic fields. A
Minco thin film heater with a nominal resistance of {50~$\Omega $ }
was attached to the bottom of the package with GE~7031 varnish in
order to improve the temperature control, and hence long term
stability required for this work.  The temperature of the vaporizer
was controlled with a LakeShore DR-91C Temperature Controller and a
calibrated CGR-1-2000 carbon-glass sensor, while the sample
temperature was controlled with a LakeShore DR-93C Temperature
Controller and a calibrated CGR-1-2000 carbon glass sensor. In order
to achieve long term temperature stability, the heater on the
vaporizer was set to attain temperature slightly below the desired
measurement point; the heater on the package was used to fine-tune and
stabilize the temperature. Thus, temperature stability of $\sim 1mK$ could be obtained and held for hours, enabling extremely detailed study of
$\tilde{Z}_s(H)|_{\omega,T}$. 

The microwave transmission amplitude $S_{21}$ of the fully assembled system
was measured using an HP~8510C Automatic Network Analyzer (ANA). All data
acquisition and computer control was carried out on an HP~745i workstation
running HP~VEE.

In order to obtain data at these low field scales it becomes essential to go
beyond simply characterizing the resonance by measuring the resonance
frequency $f_0$ and the {3~dB} bandwidth $\Delta {f}_{3~dB}$ and hence the
quality factor $Q=f_0/\Delta {f}_{3~dB}$. Thus we developed a least squares
analysis to fit our resonances to a Lorentzian form. The results of these
fits were found to agree very well with the values obtained from the
standard {3~dB} method, but provided significantly enhanced sensitivity to
small changes.

The changes in surface impedance associated with the application of
magnetic field are extracted from the measured quantities as follows: 
\begin{equation}
\begin{split}
\Delta \tilde{Z}_s(H,T) & = \tilde{Z}_s(H,T) - \tilde{Z}_s(H=0,T)\\ & = \Gamma
\left[\Delta f_{3dB}(H,T) - \Delta f_{3dB}(H=0,T) +
i2\left(f_0(0,T)-f_0(H,T)\right)\right],  
\label{Eq:ZS}
\end{split}
\end{equation}
where $\Gamma$ is a geometric factor determined from the sample
dimensions.

\section{Theory}

\subsection{Microwave vortex dynamics}

Viscoelastic models of microwave vortex dynamics, such as those developed
early on by Gittleman and Rosenblum\cite{JIGittleman66a} and later by Coffey
and Clem\cite{MWCoffey91b} consider the effects of the Lorentz force which
is exerted on the vortices by the microwave currents as the main source of
an increase in surface impedance in the presence of magnetic fields. The
applicability of these models has been validated in earlier experiments on a
variety of crystals and films \cite{JOwliaei92a,DHWu90a,SSridhar92a,SOxx96a}.

We have shown in Ref.~\cite{BAWillemsen94b} that the field induced changes
in surface impedance of a superconducting thin film can be well described
using a simple model of vortex dynamics and parameters which are consistent
with measurements on YBCO single crystals \cite{DHWu90a}, with 
\begin{equation}
\Delta \tilde{Z}_s(H,T)=\frac{i2\omega \phi _0}{\mu _0d[\kappa _s(T)+i\omega
\eta (T)]}\,{B(H)},  \label{Eq:ThinFilm}
\end{equation}
At high fields $B(H)\rightarrow \mu_0 H$, but this is clearly not the case
at low fields or in the presence of bulk pinning where the field
profile in the sample can be non-uniform. It should also be noted that
the increases in surface impedance due to penetrated flux must be
insensitive to the orientation of the penetrated vortices. Thus, it is
insufficient to simply replace $B(H)$ with an average flux density
$\overline{B(H)}\equiv \Phi /S\equiv (1/S)\int\limits_SB(H)dS$, where
$S$ is the surface of the sample perpendicular to the applied field.
The desired quantity is the average {\em absolute} flux density,
$\overline{\left| B(H)\right| }\equiv (1/S)\int\limits_S\left|
  B(H)\right| dS$. Thus, $\Delta \tilde{Z}_s(H,T)$ can be thought of
as counting the absolute number of vortices in the sample since it is
directly proportional to $\overline{\left| B(H)\right| }$. Hence for
non-uniform flux profiles, the above equation is modified to:

\begin{equation}
\Delta \tilde{Z}_s(H,T)=\frac{i2\omega \phi _0}{\mu _0d[\kappa _s(T)+i\omega
\eta (T)]}\,\overline{\left| B(H)\right| },  \label{Eq:inhomogenous}
\end{equation}

\subsection{Critical State Flux Penetration Models}

The penetration of magnetic flux into a superconductor depends
crucially on two elements, geometry and pinning. These are both
incorporated in what is usually called the critical state model,
originally due to Bean\cite {CPBean62a}. Bulk pinning is described in
terms of the critical current $J_c$, while the field profiles depend
heavily on the sample geometry. We will compare the results for the
more traditional slab geometry in a longitudinal field to those for a
thin strip in a perpendicular field

In the following we assume that $H_{c1}$ is negligible (i.e.\ that flux
penetrates for all fields), and that $J_c(B)=J_c(0)$ (sometimes referred to
as the Bean-London model). Although hysteresis loops are usually described
in terms of the magnetization $M(H)$, the average absolute induced flux
density $\overline{\left| B(H)\right| }$ is the relevant parameter to use in
our case, as described above.

\subsubsection{Bulk Geometries (Slab)}

\label{Sec:BulkBean}

For a slab of infinite extent of width $W$, the flux profiles can
easily be written down\cite{MTinkham75}, and one can immediately
compute $\overline{\left| B(H)\right|
  }=(1/W)\int\limits_{-W/2}^{W/2}\left| B(x)\right| {\rm d}x$.  When
all vortices point in the same direction, $\overline{\left| B(H)\right|
  }=\Phi /W\equiv (1/W)\int\limits_{-W/2}^{W/2}B(x){\rm d}x$, where
$\Phi $ is the induced flux per unit length in the sample. As the
applied field is increased flux penetrates gradually from the edges of
the slab, until a characteristic field $H_{s,{\rm bulk}}=J_cW/2$ is
reached and the two flux fronts meet at the center of the slab. For
increasing field applied to a slab initially in the virgin state
\begin{equation}
\overline{\left| B(H)\right| }=\begin{cases}
\frac{1}{2}\frac{{H}^2}{H_{s,{\rm bulk}}},&\text{for $ H<H_{s,{\rm bulk}}$ }\\
H-\frac{1}{2}H_{s,{\rm bulk}},&\text{for $ H>H_{s,{\rm bulk}}$ }. \end{cases}  \label{Eq:VirginSlab}
\end{equation}
When the field is reduced after reaching a maximum field $H_{{\rm max}}>{H}_{s,{\rm bulk}}
$, the field and current profiles can be constructed using a simple
construction. 
\begin{equation}
f_{\downarrow }(x,H,H_{s,{\rm bulk}},H_{{\rm max}})=f(x,H_{{\rm max}},H_{s,{\rm bulk}})-f(x,H_{{\rm max}}-H,2H_{s,{\rm bulk}}),  
\label{Eq:Construction}
\end{equation}
where $f$ can be any of the following quantities $J$, $B$ or $M$.
$f_{\downarrow }$ is that same quantity as the field is reduced from
$H_{{\rm max}}$. Thus, as the applied field is reduced, the flux
density at the edges is gradually reduced. At $H=0$ the standard
sandpile flux profile can be found at the center of the sample. As the
field is now increased in the other direction, negative vortices enter
at the edges but the positive vortices at the center remain until
$\left| H\right| >H_{s,{\rm bulk}}$. The curvature in the response
arises from the insensitivity to direction of flux, and would not be
observed in quantities such as $M(H)$ or $\Phi (H)$. The calculated
hysteresis in $\overline{\left| B(H)\right| }$ for a slab is shown in
Fig.~\ref{Fig:Slab}

\subsubsection{Thin Film in a Perpendicular Magnetic Field}

\label{Sec:PerpBean}

A number of theorists\cite{EHBrandt93a,MDarwin93a,EZeldov94a,JRClem94b} have recently
studied the problem of thin strips or disks in a perpendicular magnetic
field, using conformal mapping techniques introduced early on by Norris\cite
{WTNorris70a}. Analytic solutions for the current and flux density
distributions in the strip are thus now becoming available for this problem,
which could heretofore only be considered numerically.

In what follows we will briefly summarize the central results of Ref.~\cite
{EHBrandt93a} recast in a form which better suits our purposes. Again, as in
the case of the traditional Bean model for bulk geometries we ignore $H_{c1}$
and assume that $J_c(B)=J_c(0)$.

The geometry is that of an infinitely long strip of thickness $d$ and
width $2a$ aligned such that the sample lies in the $xy$ plane with
its width along the $\hat{y}$ axis. In the Meissner state the current
distribution must be such that $B_z(y)=0$ for $\left|y\right|<a$.

If flux can now penetrate to a distance $b<a$, the current distribution must
change. $b$ can be determined by imposing the condition that $\bar{J}(b)$ is
finite, as well as the boundary condition on $B$ ($B(x)=\mu _0H$ far away
from the sample. This leads to $b=a/\cosh \left( H/H_{s,{\rm strip}}\right) $, where $H_{s,{\rm strip}}$
is now a characteristic field related to the critical current and the
geometry, $H_{s,{\rm strip}}=J_cd/\pi $. It should be noted that unlike at $H_{s,{\rm slab}}$ in the
case of a slab, the two flux fronts do not meet at $H=H_{s,{\rm strip}}$. In fact, $b$ is
always non-zero for finite fields, although at some point $b$ will be
smaller than the effective size of a single vortex $~\lambda $ and the
validity of these expressions must be questioned. The resulting sheet
current distribution can then be written as 
\begin{equation}
K_x(y)\equiv \int\limits_{-d/2}^{d/2}J_x(z,y)=\begin{cases}
\frac{2J_cd}{\pi}\tan^{-1}\frac{y\sqrt{a^2-b^2}}{a\sqrt{b^2-y^2}}, &
\left|y\right|<b\\ J_cd\frac{y}{\left|y\right|}, & b<\left|y\right|<a.\\
\end{cases}  \label{Eq:BrandtJ1}
\end{equation}
\begin{equation}
B_z(y)=\begin{cases} 0, & \left|y\right|<b\\
\mu_0H_c\tanh^{-1}\frac{a\sqrt{y^2-b^2}}{\left|y\right|\sqrt{a^2-b^2}}, &
b<\left|y\right|<a\\
\mu_0H_c\tanh^{-1}\frac{\left|y\right|\sqrt{a^2-b^2}}{a\sqrt{y^2-b^2}}, &
a<\left|y\right| \\ \end{cases}  \label{Eq:BrandtH1}
\end{equation}
It should be noted that $B_z (y)$ has a logarithmic singularity at the edges in
this expression.

Thus, we can use Eq.~\ref{Eq:BrandtH1} to evaluate 
\begin{equation}
\begin{split}
\overline{\left| B(H)\right| }& =(1/2a)\int\limits_{-a}^aB(y){\rm d}y \\
& =\mu _0H_{s,{\rm strip}}\ln \cosh \frac H{H_{s,{\rm strip}}}
\end{split}
\label{Eq:LnCosh}
\end{equation}
Note that this has some striking similarities with the results for the slab
geometry: (i) a characteristic field scale ($H_{s,{\rm bulk}}$ or $H_{s,{\rm strip}}$) related to the
critical current density, (ii) $\overline{\left| B(H)\right| }\propto H^2$ at
low fields $H\ll H_s$, and (iii) $\overline{\left| B(H)\right| }\propto H$ at
high fields $H\gg H_s$.

Assuming that Eq.~\ref{Eq:Construction} applies just as for the slab,
the hysteretic response can be constructed with $H_{s,{\rm strip}}$ taking the place
of $H_{s,{\rm bulk}}$, and is presented in Fig.~\ref {Fig:BeanHyst}.

\section{$\tilde{Z}_s(H)$: Experimental Results}

All of the results presented here were obtained with the sample at
fixed temperature, and initially in the virgin state. This allows us
to separate out the initial penetration of flux from the overall
response and hysteresis which is observed at higher fields. The sample
was warmed up well above $T_c$ and cooled down in zero field between
runs in order to ensure that the sample was really in the virgin
state.

The data for the virgin state flux penetration from $<1$~{\rm{Oe}} to $6$~T fall
naturally into $3$ field regimes:

\begin{enumerate}
\item changes occur initially on very small field scales which we call
  $H_i(4.2K)\sim 100{\rm{Oe}}$.

\item  This is followed by the majority of the flux penetration at field
scales of around $H_s(4.2K)\sim 5000\,{\rm{Oe}}$ characterized by a superlinear
dependence of the impedance on $H$, and indicating an inhomogeneous flux
distribution.

\item At very high fields $H\gg H_s$ the dependence of
  $\tilde{Z}_s(H)$ becomes linear in $H$ indicating uniform flux
  penetration.
\end{enumerate}

\subsection{Initial Penetration of Flux}

The detailed behavior up to ${400~{\rm{Oe}}}$ at $4.2K$ is presented in Fig.~\ref
{Fig:LVirgin}, where it can clearly be seen to be characterized by a field
scale $H_i(4.2K)\sim 100{\rm{Oe}}$, and that above and below $H_i$ the impedance
appears to be approximately linear. The actual dependence is well described
by the functional form: 
\begin{equation}
\overline{\left| B(H)\right| }=C_i\tanh (H/H_i)+C_lH  \label{NU:LEff}
\end{equation}
and is shown in the figure as a solid line. The value of $H_i$ used in the
fit is indicated by an arrow.

It is easy to extract $H_i$ from the data by a linear extrapolation of
the behavior above and below the bend and is presented in
Fig.~\ref{Fig:Hi}. $H_i $ is chosen to be the intersection point of
these two lines. This is nearly equivalent to the field scale $H_i$
that enters in Eq.~\ref{NU:LEff}, but does not presuppose any specific
functional dependence of $\tilde{Z}_s(H) $ at low fields.

Is this low field scale simply a manifestation of $H_{c1}$? $H_{c1}$
is typically observed as a sharp break from practically zero in
microwave measurements, due to the logarithmic singularity in $B(H)$
near $H_{c1}$.  This is not what is seen in Fig.~\ref{Fig:LVirgin}.
However, the magnetic field $H_{c1,{\rm strip}}$ at which flux
initially penetrates into a thin strip in a perpendicular magnetic
field will in general be lower than the bulk $H_{c1}$ of the material,
due to demagnetization effects. Using standard ellipsoidal
approximations to a thin strip yields an estimate of $H_{c1,{\rm
    strip}}=H_{c1}(d/2a)$.  Recent analytical
calculations\cite{EZeldov94b} find that the effect is actually smaller
than that predicted by the ellipsoidal approximation by a factor of
$\sqrt{d/2a}$, so that $ H_{c1,{\rm strip}}=H_{c1}\sqrt{d/2a}$. For
a ${\rm {0.5~\mu {m}}}\times {\rm {100~\mu {m}}}$ strip, typical of
those used for these experiments, one finds that $H_{c1,{\rm
    strip}}\sim H_{c1}/10$.

Bulk $H_{c1}$ for single crystalline YBCO was measured by Wu {\it et
  al.}\cite{DHWu90a} where they found that
$H_{c1,\parallel}(0)={\text{850}}\pm {\text{40~{\rm{Oe}}}}$, leaving
us with an upper bound on the penetration field $H_{c1,{\rm
    strip}}\sim{\text{85}}\pm {\text{4~{\rm{Oe}}}}$ for our samples.
This magnitude is comparable to the observed low temperature magnitude
of $H_i$.

If $H_i$ is simply a manifestation of $H_{c1}$, the linear field dependence
below $H_i$ is somewhat unconventional. For a conventional GL
superconductor, the effects of pairbreaking are expected to have a quadratic
field dependence below $H_{c1}$\cite{ABPippard50,BDJosephson74}. This quadratic dependence has also
been observed experimentally in radio frequency penetration depth
measurements on polycrystalline samples\cite{DHWu88a} and single crystals of
YBCO\cite{DHWu90a} for fields applied parallel the the $ab$ plane. It should
also be mentioned that Yip and Sauls\cite{SKYip92a} have shown that for a $d$
-wave superconductor, the penetration depth (and hence also the surface
resistance) should rise linearly in the Meissner state.


\subsubsection{$H_{c1}$ and Weak Links}

A quadratic field dependence is also expected from the weakly
coupled-grain model\cite{TLHylton88a,PPNguyen93a}, which has been used
to explain the functional dependence of $\tilde{Z}_s$ on low microwave
fields. Certain theories for granular superconductors do predict an
initial linear field dependence\cite{JHalbritter92a,AMPortis91a} by
considering a Bean-like model applied to the inter-granular weak
links. Weak links are proposed as an explanation for a similar initial
linear rise followed by a saturation reported in the
literature\cite{MGolosovsky96a}. We looked for, but did not find any
direct correlation between film granularity and the magnitude $H_i$.
Films of very different granularity (as evidenced by the rate at which
the films etched) had remarkably similar magnitudes of $H_i$. Thus,
the linear dependence on field cannot conclusively be attributed to
the presence of weak links in the samples.

\subsection{Scaling of the data for moderate fields}

The field induced change in surface impedance, $\Delta
\tilde{Z}_s(H,T,\omega )=\tilde{Z}_s(H,T,\omega
)-\tilde{Z}_s(H\rightarrow 0,T,\omega )$ at moderate fields
($H_i<H<{\text{4~T}}$) scales, giving a field scale $H_s$ and an
impedance scale which can be determined from the slope at high fields.

The scaled data over this broad field range are well described by the
following expression 
\begin{equation}
\overline{\left| B(H) \right|} =C_s\ln \cosh (H/H_s)  \label{NU:Eff}
\end{equation}
where $H_s$ is the field scale well above which the flux has largely penetrated
throughout the sample and $B(H)\approx \mu _0H$.

The scaling function used above is exactly that obtained from the critical
state model of rectangular strips presented in Sec.~\ref{Sec:PerpBean}. This
would imply that $J_c$ must depend only weakly on $H$, and only one field
scale (related to $J_c(T)$) determines the response. Thus, we should now be
able to estimate a critical current density $J_c$ from the experimentally
observed field scale and the film thickness using $H_s=J_cd/\pi $.

A similar scaling was also observed in data taken from a ring
resonator. Calculations of $\overline{\left|B(H)\right|}$ for the case
a superconducting disk using the results of Ref.~\cite{JRClem94b}
cannot be distinguished from the $\ln\cosh(H/H_{s,{\rm strip}})$
behavior of the strip within experimental uncertainty, although the
relation of the scaling field $H_{s,{\rm disk}}$ to $J_c$ is
different. We have presented the temperature dependence of $J_c(T)$ in
Fig.~\ref {Fig:ScaleJc} for the ring resonator, obtained using
Eq.\ref{NU:Eff}. (This overestimates the true $J_c$ values by a factor
of up to $\pi /2$, due to the ring geometry although the magnitudes
are consistent).

When combined with the $J_c$ values estimated from the nonlinear
response at high $T$ on the same resonator from
Ref.~\cite{BAWillemsen95a} it would appear that the two estimates are
consistent, and can well be described overall by a $(1-t^2)^2$
dependence reminiscent of a mean field thermodynamic critical field,
$H_c$. Also the magnitudes obtained, $J_c({\rm {77~K})\approx 2\times
  {10^6~A/cm^2}}$, comparable with independent DC measurements on
films of the same batch.

To summarize, flux penetration into the virgin state for moderate fields is
very well characterized quantitatively by the critical state model for
perpendicular geometry.

\subsection{High field response for $H\gg H_s$}

For very high fields much greater than the self field $H_s$ the impedance
becomes linear in the applied field. In this regime the slopes can be used
to extract vortex parameters such as the pinning force constants and the
viscosity as we have done in Ref.~\cite{BAWillemsen94b}.

\section{Hysteresis}

So far we have only examined what happens as the externally applied magnetic
field is increased adiabatically. The response as the field is slowly
reduced and also reversed will be discussed in this section.

\subsection{Moderate Field Hysteresis}

The typical response when the applied field is ramped up to a field
significantly higher than $H_s$ and then reduced is presented in Fig.~\ref
{Fig:HiHyst}. As can immediately be seen this response is qualitatively
different from that predicted by the critical state model alone (presented
in Fig.~\ref{Fig:BeanHyst}). The loop has the opposite handedness, and is
much smaller than $H_s$ in width. While the reduced width of the loop may be
consistent with the presence of a geometrical or other barrier at the strip
edge, the handedness however cannot be so easily explained.

There are a number of possible ways to explain this behavior:

\begin{itemize}
\item  The relationship between $\tilde{Z}_s$ and $\overline{\left| B(H)\right| 
}$ does not apply,

\item The construction (Eq.~\ref{Eq:Construction}) used to determine
  $B_{\downarrow }(H)$ does not apply, or

\item  A fraction of the flux, principally at the edges, is sufficiently
weakly pinned to be able to leave the sample quickly as the applied field is
reduced.
\end{itemize}

\subsection{Hysteresis at Very Low Field}

In order to shed some light on the issue, we have also carried out
measurements of the hysteretic response for fields $\sim H_i$. Typical
response when the applied field is ramped up to a field $H_{{\rm
    max}}$ comparable to $H_i$ and then down to $-H_{{\rm max}}$ and
finally increased back to $H_{{\rm max}}$ is presented in
Fig.~\ref{Fig:LoHyst}. Although only $\Delta X_s(H)$ is presented in
the figure, $\Delta R_s(H)$ has qualitatively the same behavior. At
sufficiently high $T$, response qualitatively similar to that observed
in the previous section is recovered even for fields
$\left|H\right|<{\text{400~{\rm{Oe}}}}$.

To put these effects into perspective Fig.~\ref{Fig:HystComp} presents the
data of Fig.~\ref{Fig:HiHyst}, along with low field data obtained at the
same temperature. Note that the magnitude of these effects is very small,
and requires extreme care to be observed.

Clearly the clockwise nature of the hysteresis initially observed at
high fields is preserved even at these low field scales, and the drop
in $\tilde{Z}_s$ as the applied field is reduced appears to be rather
sudden, taking place within $<\text{10~{\rm{Oe}}}$.

Below we summarize the essential features of the hysteretic response:

\begin{itemize}
\item  Sudden drop in both $R_s$ and $X_s$ as the applied field is reduced,
below even the virgin state response

\item  After $H_{{\rm max}}$ is reached, the response is clearly symmetric
in $H$ about $H=0$.

\item  $\tilde{Z}_s$ at $H=0$ does not return to its initial value in the
normal state.

\item  a cusp-like feature is observed as $H$ is increased in the opposite
direction after reaching $H=0$.
\end{itemize}

A cusp-like feature is expected in magnetization as the field crosses $H=0$
in the presence of a geometrical barrier\cite{EZeldov94a}. One must also
keep in mind that the time taken to manually switch the current leads (to
reverse the field) was finite in these experiments so that any time
dependence (such as flux creep) may also be a possible source for the
observed cusp.

\subsection{Comparison to DC Magnetization}

Hysteresis loops from DC magnetization measurements on thin films and bulk
polycrystalline samples in high fields ($H>\sim1$~T) generally appear quite
different from those we report here, and are often more consistent with
critical state models. $M(H)$ loops are generally counter clockwise and
estimates of $J_c(H)$ are routinely extracted from the directly from the
loop width. Measurements on single crystalline samples also have the
``correct'' handedness although the fishtail effect is observed in the vast
majority of samples.

This is in contrast to the results presented here where the hysteresis loops
have the opposite handedness are much narrower than expected from the
critical state model. In fact, rather than being related to $H_s$, the
hysteresis width appears to be governed by a field scale on the order of $H_i
$.

\section{Conclusion}

The linear electrodynamic response of vortices in thin films in the
perpendicular geometry has been investigated within the framework of
recent theoretical work. While flux entry into a sample initially in
the virgin state appears to be well described by the critical state
model, flux exit when the field is reversed does not follow the model. The expected hysteresis from the critical state model (of magnitude $\approx 2H_s$), is {\em \ 
  not} observed. In fact, the observed behavior would imply that the
absolute flux density $\overline{\left| B(H)\right| }$ drops suddenly as
the applied magnetic field is reduced.  The presence of geometrical
barriers at the strip edges can help to reduce the level of hysteresis
expected, but will not lead to the observed response.

\section*{Acknowledgements}

This work was supported by Rome Laboratory and the NSF through
NSF-DMR-9223850 and NSF-DMR-9623720. The authors thank R.~S. Markiewicz and D.P.Choudhury for useful discussions and Jos\'e Silva for his technical assistance. Balam A. Willemsen is currently at Superconductor Technologies, Inc., Santa
Barbara, CA.

\bibliographystyle{prsty}
\bibliography{strings,big,sc96}

\begin{thebibliography}{10}

\bibitem{AForkl93a}
A. Forkl, Phys. Scr. {\bf T49},  148  (1993).

\bibitem{BAWillemsen94b}
B.~A. Willemsen, J.~S. Derov, J.~H. Silva, and S. Sridhar, Appl. Phys. Lett.
  {\bf 67},  551  (1995).

\bibitem{BAWillemsen95a}
B.~A. Willemsen, J.~S. Derov, J.~H. Silva, and S. Sridhar, IEEE Trans. Appl.
  Supercond. {\bf 5},  1753  (1995).

\bibitem{JIGittleman66a}
J.~I. Gittleman and B. Rosenblum, Phys. Rev. Lett. {\bf 16},  734  (1966).

\bibitem{MWCoffey91b}
M.~W. Coffey and J.~R. Clem, Phys. Rev. Lett. {\bf 67},  386  (1991).

\bibitem{JOwliaei92a}
J. Owliaei, S. Sridhar, and J. Talvacchio, Phys. Rev. Lett. {\bf 69},  3366
  (1992).

\bibitem{DHWu90a}
D.-H. Wu and S. Sridhar, Phys. Rev. Lett. {\bf 65},  2074  (1990).

\bibitem{SSridhar92a}
S. Sridhar {\it et~al.}, Phys. Rev. Lett. {\bf 68},  2220  (1992).

\bibitem{SOxx96a}
S. Oxx {\it et~al.}, Physica C {\bf 264},  103  (1996).

\bibitem{CPBean62a}
C.~P. Bean, Phys. Rev. Lett. {\bf 8},  250  (1962).

\bibitem{MTinkham75}
M. Tinkham, {\em Introduction to Superconductivity} (McGraw-Hill, Inc., New
  York, 1975).

\bibitem{EHBrandt93a}
E.~H. Brandt and M. Indenbom, Phys. Rev. B {\bf 48},  12893  (1993).

\bibitem{MDarwin93a}
M. Darwin {\it et~al.}, Phys. Rev. B {\bf 48},  13192  (1993).

\bibitem{EZeldov94a}
E. Zeldov {\it et~al.}, Phys. Rev. Lett. {\bf 73},  1428  (1994).

\bibitem{JRClem94b}
J.~R. Clem and A. Sanchez, Phys. Rev. B {\bf 50},  9355  (1994).

\bibitem{WTNorris70a}
W.~T. Norris, J. Phys. D {\bf 3},  489  (1970).

\bibitem{EZeldov94b}
E. Zeldov {\it et~al.}, Physica C {\bf 235-240},  2761  (1994), (Proc. M2S HTSC
  IV).

\bibitem{ABPippard50}
A.~B. Pippard, Proc. Roy. Soc. London {\bf Ser A 203},  210  (1950).

\bibitem{BDJosephson74}
B.~D. Josephson, J. Phys. F {\bf 4},  751  (1974).

\bibitem{DHWu88a}
D.-H. Wu, C.~A. Shiffman, and S. Sridhar, Phys. Rev. B {\bf 38},  R9311
  (1988), rapid Comm.

\bibitem{SKYip92a}
S.~K. Yip and J.~A. Sauls, Phys. Rev. Lett. {\bf 69},  2264  (1992).

\bibitem{TLHylton88a}
T.~L. Hylton {\it et~al.}, Appl. Phys. Lett. {\bf 53},  1343  (1988).

\bibitem{PPNguyen93a}
P.~P. Nguyen, D.~E. Oates, G. Dresselhaus, and M.~S. Dresselhaus, Phys. Rev. B
  {\bf 48},  6400  (1993).

\bibitem{JHalbritter92a}
J. Halbritter, J. Appl. Phys. {\bf 71},  339  (92).

\bibitem{AMPortis91a}
A.~M. Portis {\it et~al.}, Appl. Phys. Lett. {\bf 58},  307  (1991).

\bibitem{MGolosovsky96a}
M. Golosovsky {\it et~al.}, Supercond. Sci. Tech. {\bf 9},  1  (1996).

\end{thebibliography}

\mediumtext
  \begin{table}[htb]
    \begin{center}
      \leavevmode
      \begin{tabular}{cccccc}
        Resonator&Geometry&Line Width&Resonator Length& Film Thickness & Substrate Thickness\\ \tableline
        N4& Straight & 100~$\mu$m & 1.32~cm& 0.38~$\mu$m& 250~$\mu$m  \\
        N3& Ring &  100~$\mu$m & 2.00~cm& 0.375~$\mu$m&500~$\mu$m\\
      \end{tabular}
      \caption{The two resonators used for this work}
      \label{tab:Resonators}
    \end{center}
  \end{table}

\widetext


\begin{figure}[htb]
 \begin{center}
  \leavevmode
  \caption{Typical virgin state response: Raw $ \Delta f$ data
   for increasing applied field for a sample initially in the
   virgin state for resonator N3 at
   {3.7~GHz} and {5~K}. The dashed line is the
   expression of Eq.~\protect\ref{NU:Eff}. The arrows indicate the
   two field scales, $H_i$ and $H_s$.}
  \label{Fig:Virgin}
 \end{center}
\end{figure}

\begin{figure}[htb]
 \begin{center}
  \leavevmode
  \caption{The calculated hysteresis in $ \overline{\left|B(H)\right|}$ 
    for a thick slab in a field applied parallel to its surface.  Both
    axes are in units of $ H_{s,{\rm bulk}}$, with $
    H_{\mathrm{max}}=10H_{s,{\rm bulk}}$.}
  \label{Fig:Slab}
 \end{center}
\end{figure}

\begin{figure}[htb]
 \begin{center}
  \leavevmode
 \end{center}
 \caption{Hysteretic response in $ \overline{\left|B(H)\right|}$ 
  for a thin strip in a perpendicular field calculated numerically
  from the field profiles discussed in the text. Both axes are in
  units of $ H_{s,{\rm strip}}$, and $ H_{\mathrm{max}}=10H_{s,{\rm strip}}$.}
 \label{Fig:BeanHyst}
\end{figure}

\begin{figure}[htb]
 \begin{center}
  \leavevmode
  \caption{Typical virgin state response: Raw $ \Delta f$ data
   for increasing applied field for a sample initially in the
   virgin state for resonator N4 at
   {3.7~GHz} and {5~K}. The dashed line is the
   expression of Eq.~\protect\ref{NU:LEff}. The arrow indicates the
   field scale described in the text.}
  \label{Fig:LVirgin}
 \end{center}
\end{figure}

\begin{figure}[htb]
 \begin{center}
  \leavevmode
  \caption{Temperature dependence of the initial penetration field scale 
    for resonator N4.  The straight dashed line is a least squares fit
    to the data. The dotted line presents a BCS temperature dependence
    for $ H_{c1}$.}
  \label{Fig:Hi}
 \end{center}
\end{figure}

\begin{figure}[htb]
 \begin{center}
  \leavevmode
 \end{center}
 \caption{%
  Raw (left) and scaled (right) plot of the field induced changes is
  surface resistance. Data presented is for resonator N3 at
  {3.7~GHz} and $ T=$ (5.0, 15.0, 20.0, 25.0,
  31.0, 40.0 and {50.0~K}). The slope increases with
  increasing $ T$ temperatures. The solid line represents the
  expected behavior for a strip. A dashed line representing the
  expected behavior for a disk is also presented, but is
  indistinguishable from the solid line.}
 \label{Fig:TScale}
\end{figure}

\begin{figure}[htb]
 \begin{center}
  \leavevmode
 \end{center}
 \caption{Field scale $ H_s$ for resonator N3 at
  {3.7~GHz}. The dashed and dotted lines are $ (1-t^2)$ 
  and $ (1-t^2)^2$ dependences respectively}
 \label{Fig:ScaleJc}
\end{figure}

\begin{figure}[htb]
 \begin{center}
  \leavevmode
  \caption{Hysteretic response of $ R_s(H)$ for $ H_{\mathrm{max}}\gg H_s$.
   Results are presented for resonator N4 at {3.7~GHz} and {10~K}.}
  \label{Fig:HiHyst}
 \end{center}
\end{figure}

\begin{figure}[htb]
 \begin{center}
  \leavevmode
  \caption{Hysteretic response of $ X_s(H)$ for $ H_{\mathrm{max}}\sim H_i$.
   Results are presented for resonator N4 at {3.7~GHz} and {35~K}.}
  \label{Fig:LoHyst}
 \end{center}
\end{figure}
\begin{figure}[htb]
 \begin{center}
  \leavevmode
 \end{center}
 \caption{Low field hysteresis in $ R_s(H)$ at high temperatures.
  Results are presented for resonator N4 at {3.7~GHz} and
  {84~K}.}
 \label{fig:LoHystHiT}
\end{figure}

\begin{figure}[htb]
  \begin{center}
    \leavevmode
  \end{center}
  \caption{Magnified view of the hysteresis data for $ R_s$ presented in Fig.~\protect\ref{Fig:HiHyst} for moderate fields and low fields (inset).}
  \label{Fig:HystComp}
\end{figure}


  \includegraphics*[angle=90,width=\textwidth]{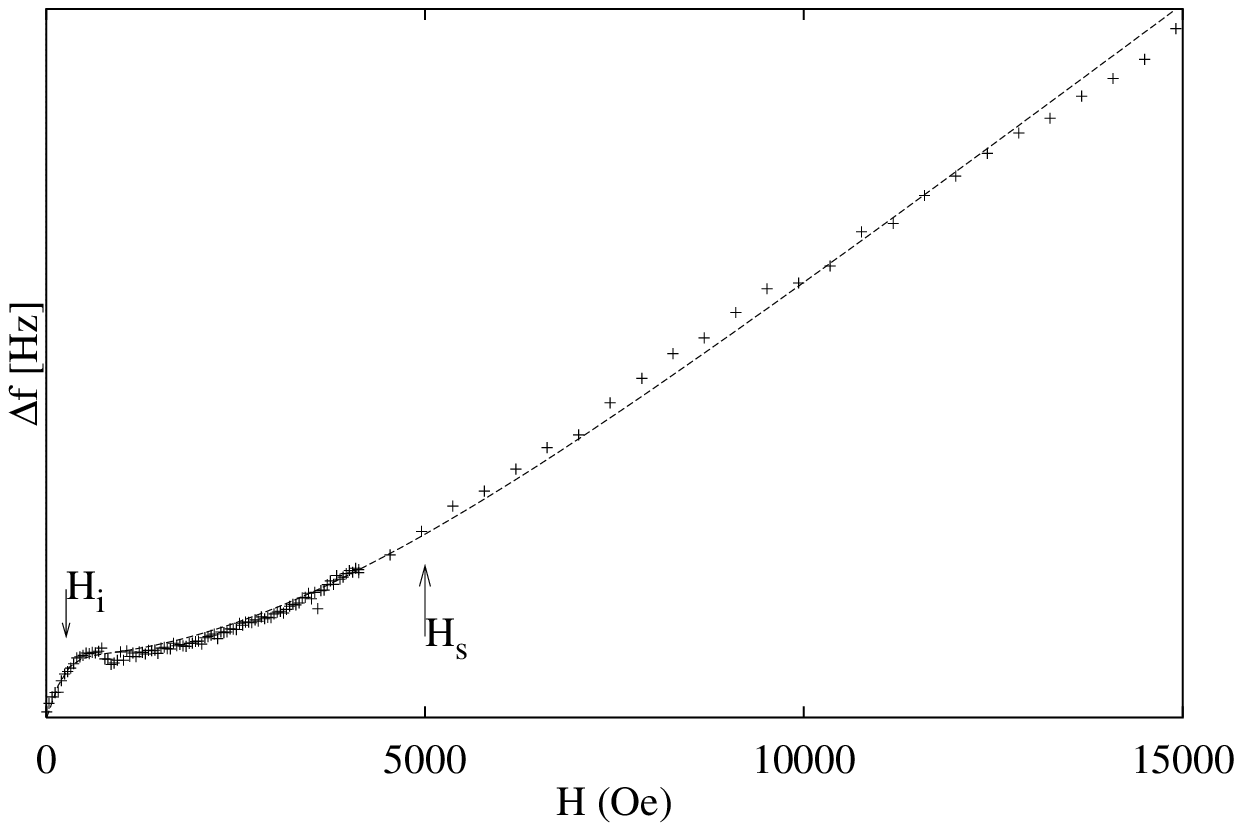}
  \newpage
  \includegraphics*[angle=90,width=\textwidth]{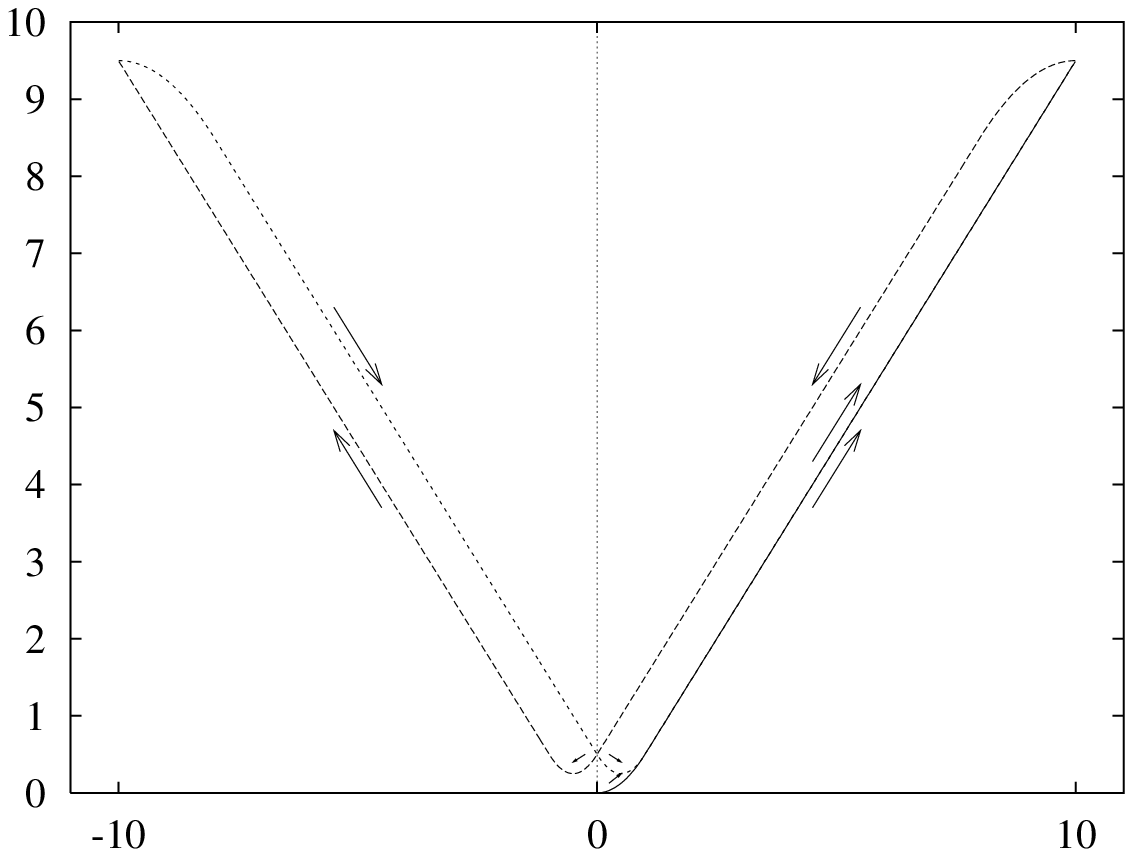}
  \newpage
  \includegraphics*[angle=90,width=\textwidth]{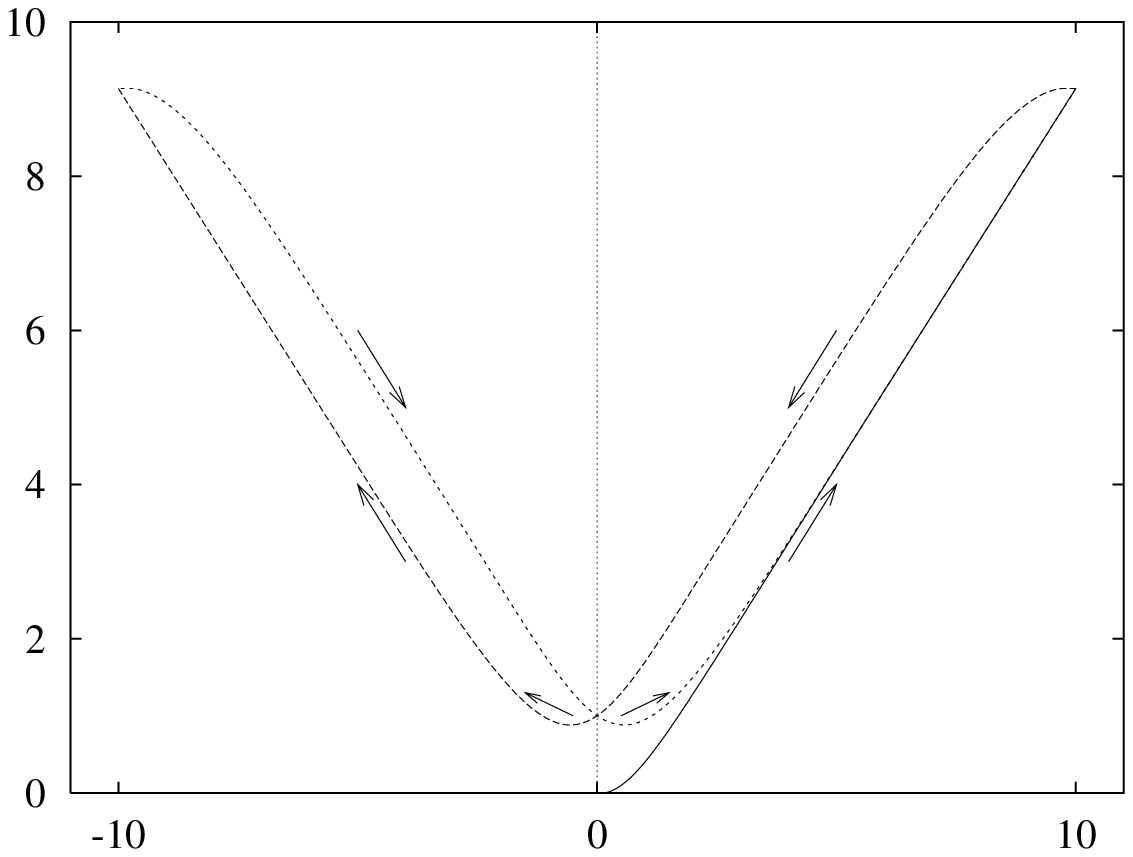}
  \newpage
  \includegraphics*[angle=90,width=\textwidth]{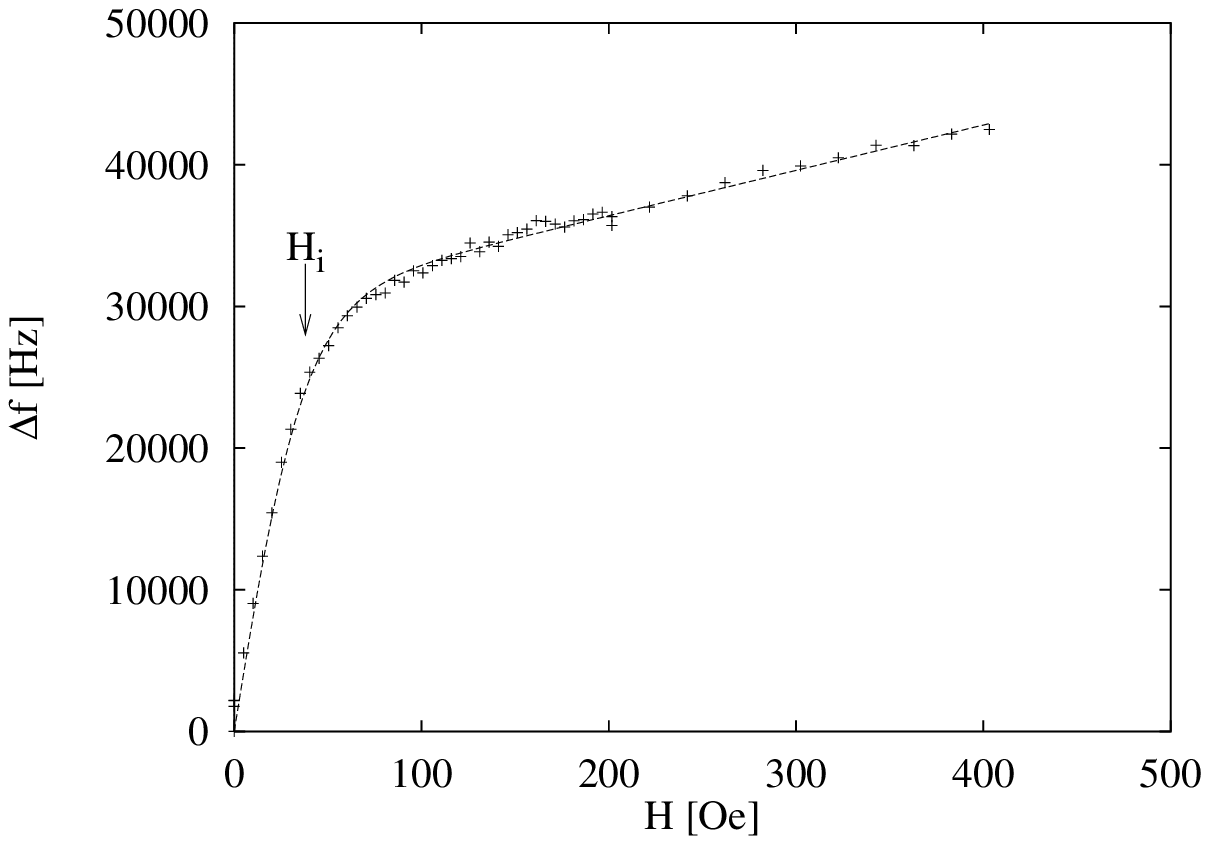}
  \newpage
  \includegraphics*[angle=90,width=\textwidth]{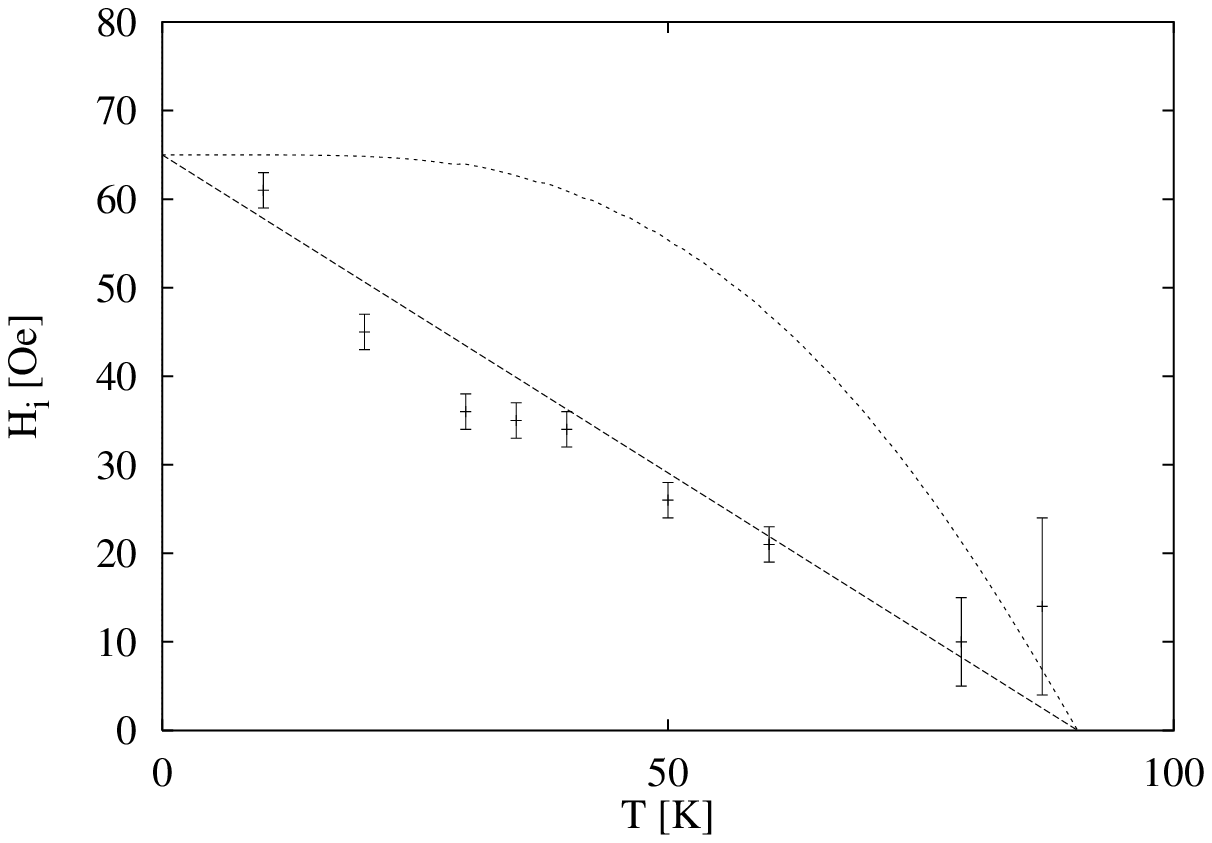}
  \newpage
  \includegraphics*[angle=90,width=\textwidth]{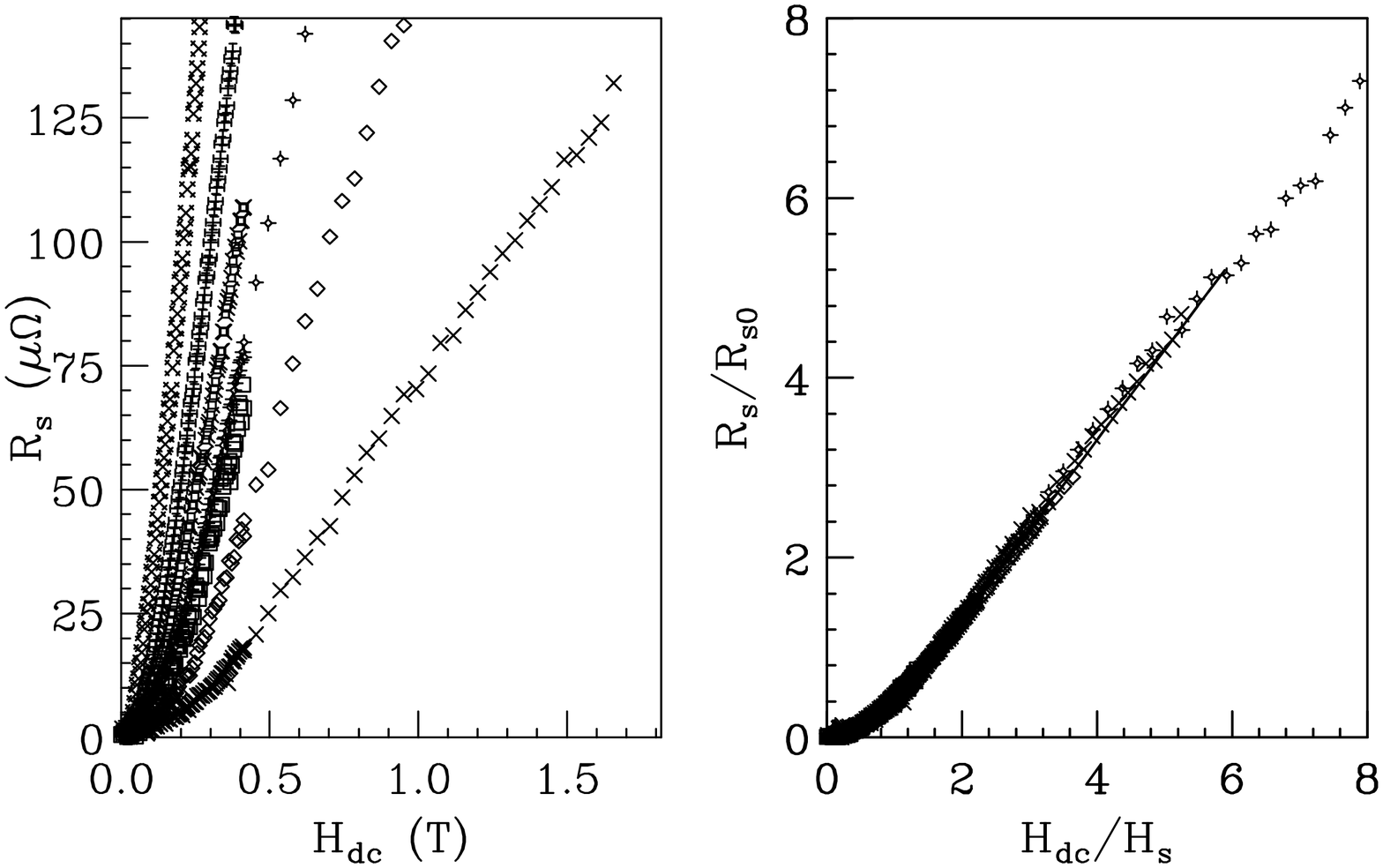}
  \newpage
  \includegraphics*[angle=90,width=\textwidth]{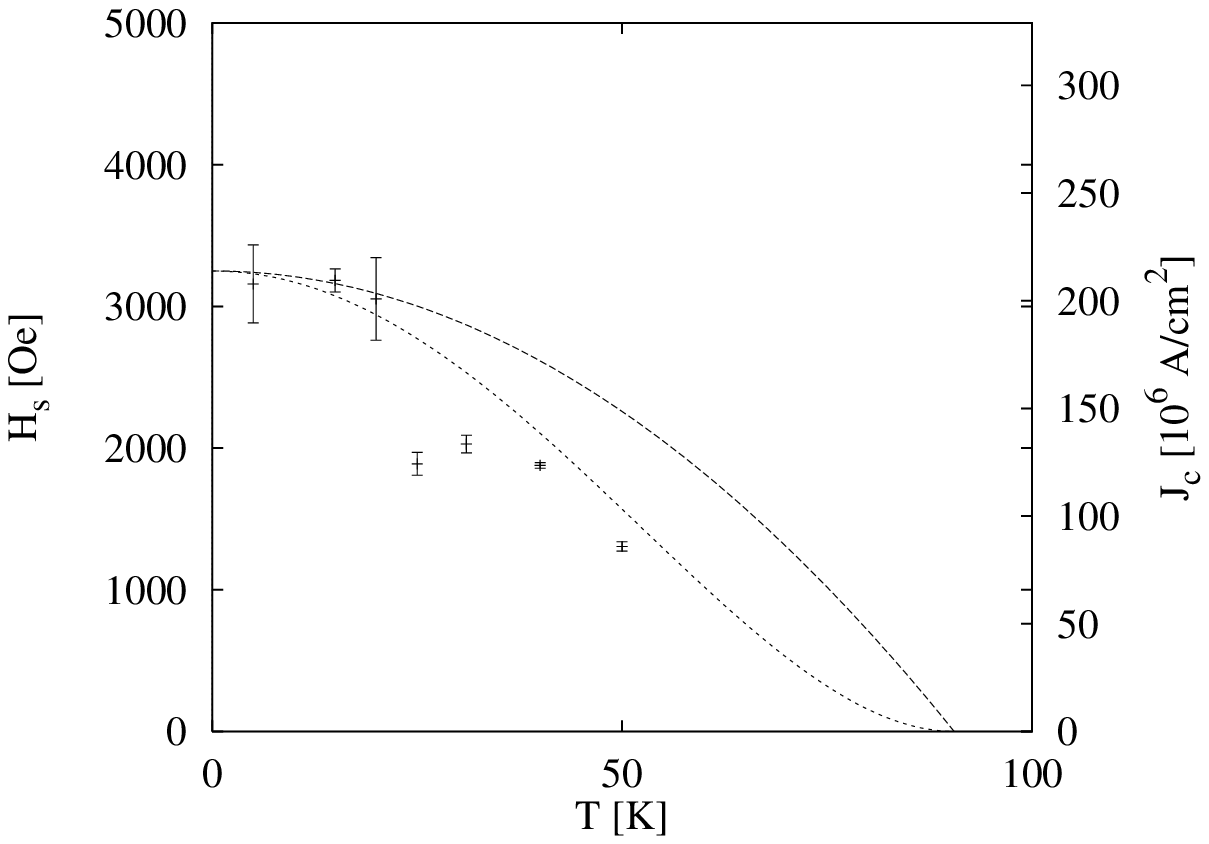}
  \newpage
  \includegraphics*[angle=90,width=\textwidth]{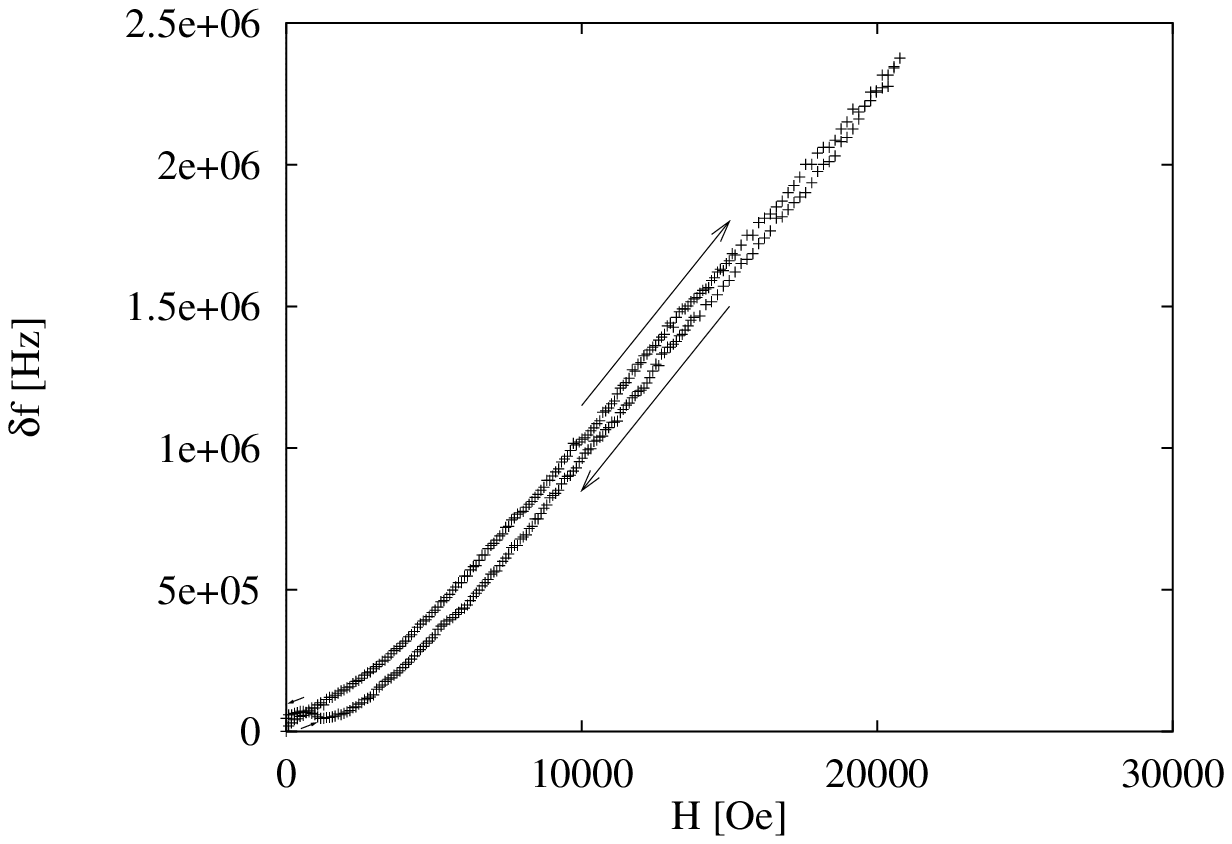}
  \newpage
  \includegraphics*[angle=90,width=\textwidth]{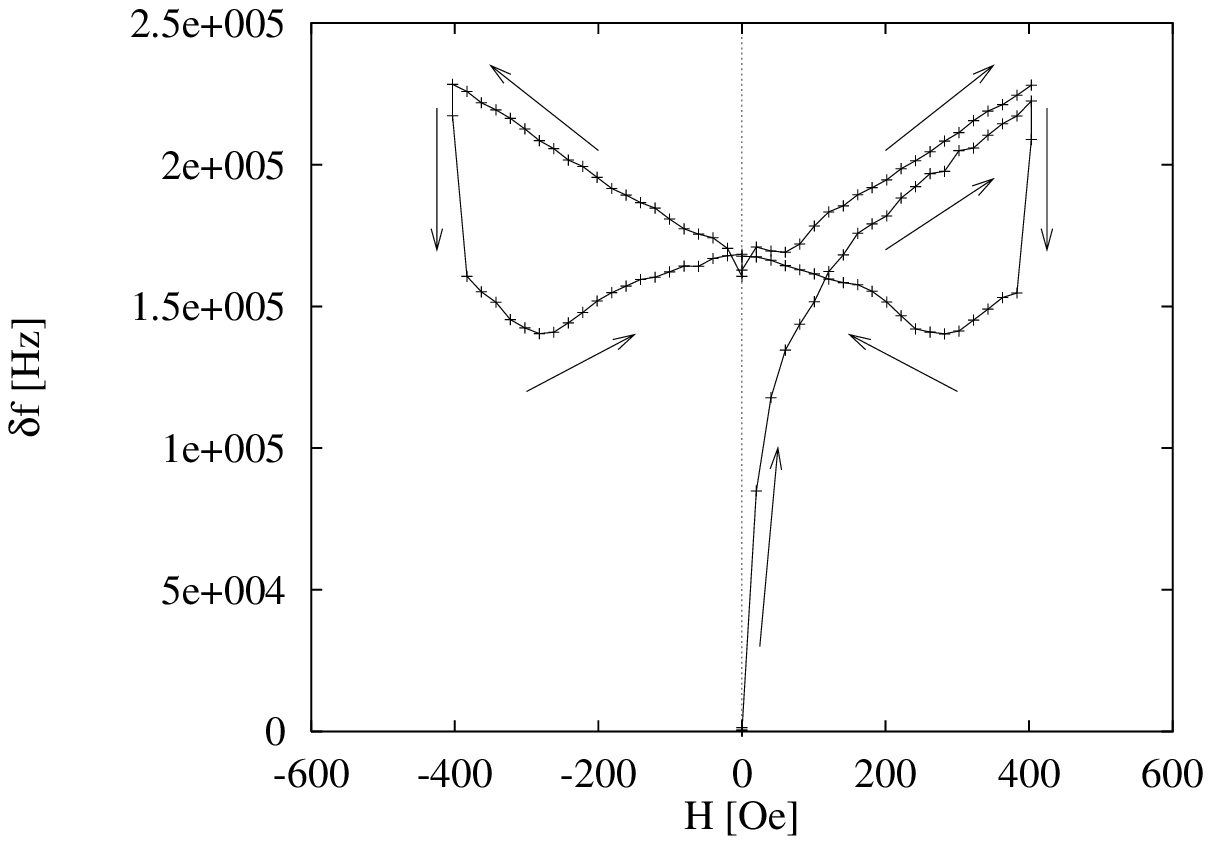}
  \newpage
  \includegraphics*[angle=90,width=\textwidth]{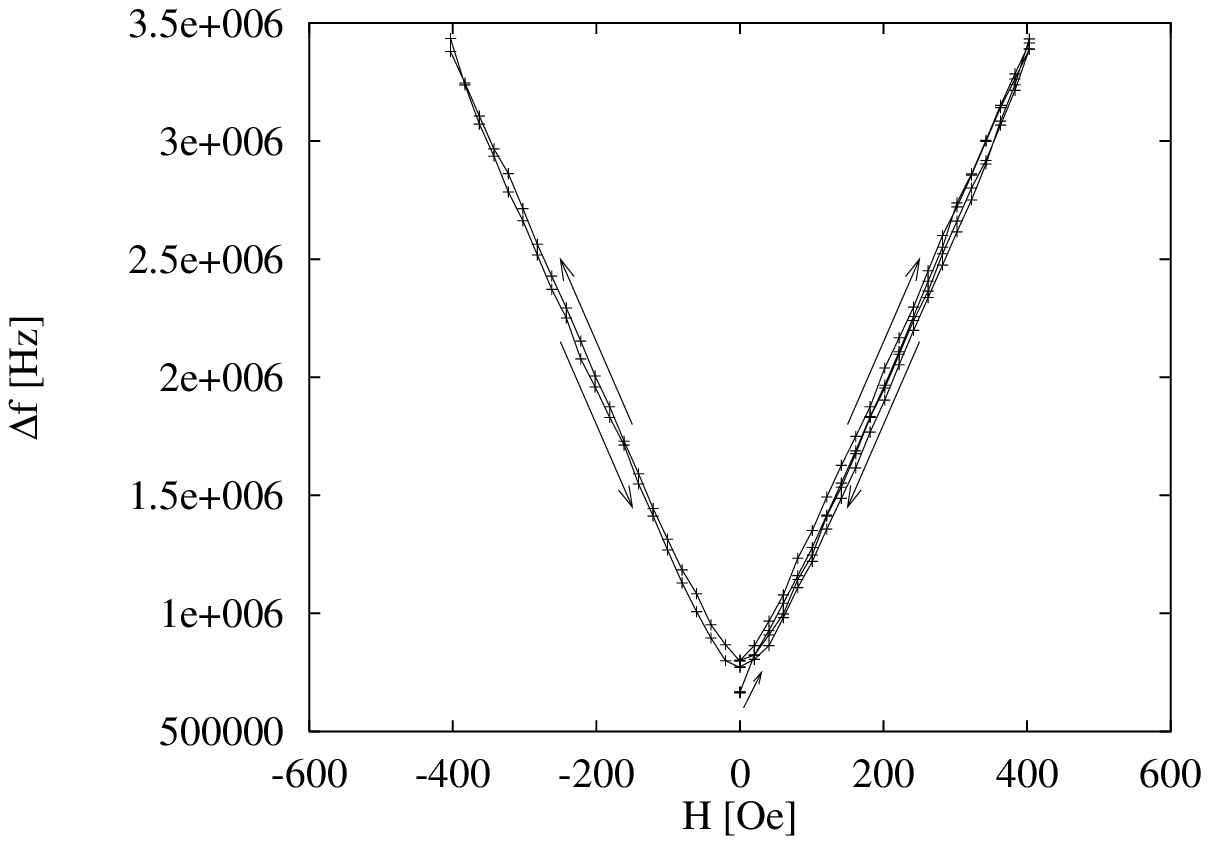}
  \newpage
  \includegraphics*[angle=90,width=\textwidth]{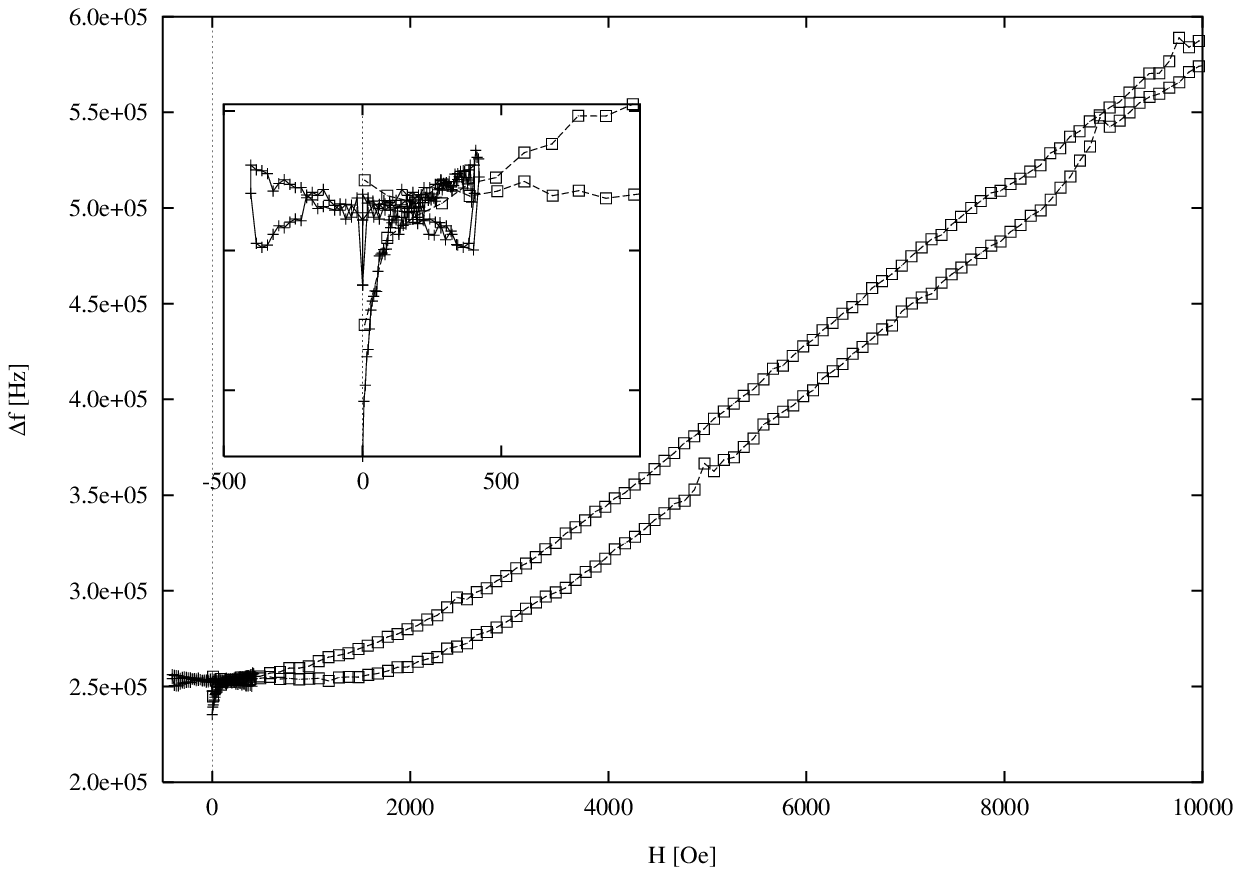}

\end{document}